\documentclass[12pt]{article}
\newcommand{\id}{i\!\!\not\!\partial}
\newcommand{\as}{\not\!\! A}

\newcommand{\D}{{\cal{D}}}
\newcommand{\be}{\begin{equation}}
\newcommand{\ee}{\end{equation}}
\newcommand{\ba}{\begin{eqnarray}}
\newcommand{\ea}{\end{eqnarray}}
\begin{document}
\title{The Wess-Zumino-Witten term in non-commutative
two-dimensional fermion models}
\author{%
E.F.~Moreno\thanks{Investigador CONICET} ~and
F.A.~Schaposnik\thanks{Investigador CICBA}\\ {\normalsize\it
Departamento de F\'{\i}sica, Universidad Nacional de La Plata}\\
{\normalsize\it
  C.C.~67, (1900) La Plata, Argentina}%
}
\date{\hfill}
\maketitle
\begin{abstract}
We study the effective action associated to the Dirac operator in
two dimensional non-commutative Field Theory. Starting from the
axial anomaly, we compute the determinant of the Dirac operator
and we find that even in the $U(1)$ theory, a Wess-Zumino-Witten
like term arises.
\end{abstract}
\date{}
\maketitle
\newpage
\section{Introduction}

Interest in non-commutative spaces has been renewed after the
discovery that non-commutative gauge theories naturally arise when
D-branes with constant B fields are considered
\cite{cds}-\cite{dh}. These works as well as that in \cite{sw}
prompted many investigations both in field theory and in string
theory (see references in \cite{sw}). Concerning gauge field
theories, recent results on chiral and gauge anomalies
\cite{as}-\cite{bst} have shown that well-known results on
``ordinary'' models extend naturally and interestingly to the case
in which non-commutative spaces are considered. In this work we
consider a problem which can be seen as closely related to that of
anomalies, namely the evaluation of the two-dimensional fermion
determinant in non-commutative space-time. This problem is of
interest not only for the analysis of two-dimensional QED and QCD
in non-commutative space, but also in connection with abelian and
non-Abelian bosonization since, as it is well-known, the knowledge
of the fermion determinant leads more or less directly to the
bosonization rules.

We start by evaluating in Section II the chiral anomaly in
two-dimensional non-commutative space-time in  a way adapted to the
calculation of fermion determinants through integration of the
anomaly. This last is done in Section III where both the Abelian
and ($U(N)$) non-Abelian fermion determinant is calculated
exactly. In both cases we obtain for the determinant a
Wess-Zumino-Witten term. Consequences of our results and possible
extensions are discussed in section IV.

\section{The chiral anomaly}

\subsection*{Conventions}
As usual, we define the $*$-product between a pair of functions
$\phi(x)$, $\chi(x)$ as
\begin{eqnarray}
\phi*\chi(x) &\equiv& \exp\left( \frac{i}{2} \theta^{\mu\nu}
\partial_{x_\mu}\partial_{y_\nu} \right) \left.
 \phi(x)\chi(y)\right\vert_{x=y}
\nonumber\\ &=& \phi(x)\chi(x) + \frac{i}{2} \theta^{\mu\nu}
\partial_\mu\phi \partial_\nu \chi(x)
+ O(\theta^2)\; , \label{2}
\end{eqnarray}
and the (Moyal) bracket in the form
\begin{equation}
\{\phi,\chi\}(x) \equiv \phi(x) *\chi(x) - \chi(x) * \phi(x)\; ,
\label{4}
\end{equation}
so that, when applied to (Euclidean) space-time coordinates
$x^\mu, x^\nu$, one has
\begin{equation}
\{x^\mu,x^\nu\} = i \theta^{\mu\nu} \label{1}
\end{equation}
which is why one refers to non-commutative spaces. Here
$\theta^{\mu \nu}$ is a real, anti-symmetric constant tensor.
Since we shall be interested in two dimensional space-time, one
necessarily has $\theta^{\mu \nu} = \theta\; \varepsilon^{\mu
\nu}$ with $\varepsilon^{\mu \nu}$ the completely anti-symmetric
tensor and $\theta$ a real constant. In the context of string
theory, non-commutative spaces are believed to be relevant to the
quantization of D-branes in background Neveu-Schwarz constant
B-field $B_{\mu \nu}$ \cite{cds}-\cite{sw}. In this context
$\theta^{\mu \nu}$ is related to the inverse of $B^{\mu \nu}$.
Afterwards, this original interest was extended to the analysis of
field theories in non-commutative space and then, as signaled in
\cite{bst} it becomes relevant to know to what extent old problems
and solutions in standard field theory fit in the new
non-commutative framework.

A ``non-commutative gauge theory'' is defined just by using the
$*$-product each time the gauge fields have to be multiplied.
Then, even in the $U(1)$ Abelian case, the curvature $F_{\mu\nu}$
has a non-linear term (with the same origin as the usual
commutator in non-Abelian gauge theories in  ordinary space)

\begin{eqnarray}
F_{\mu\nu} &=& \partial_\mu A_\nu - \partial_\nu A_\mu - ie
\left(A_\mu * A_\nu - A_\nu * A_\mu\right) \nonumber\\ &=&
\partial_\mu A_\nu - \partial_\nu A_\mu -
ie\{A_\mu,A_\nu\}\; .
\label{5}
\end{eqnarray}
This field strength is gauge-covariant (not gauge-invariant, even
in the Abel\-ian case) under gauge transformations which should be
represented by $U$ elements of the form
\begin{equation}
U(x) = \exp_*(i\lambda) \equiv 1 + i \lambda - \frac{1}{2}
\lambda*\lambda + \ldots \label{6}
\end{equation}
The covariant derivative implementing infinitesimal gauge
transformations takes the form
\begin{equation}
{\cal D}_\mu[A] \lambda = \partial_\mu \lambda + ie \left( \lambda
*A_\mu - A_\mu*\lambda\right) \label{7}
\end{equation}
so that an infinitesimal gauge transformation on $A_\mu$ reads as
usual
\begin{equation}
\delta A_\mu = \frac{1}{e} {\cal D}_\mu \lambda \label{in}
\end{equation}
Concerning finite gauge transformations, one has
\begin{equation}
A_\mu^{U} = \frac{i}{e} U(x) * \partial_\mu U^{-1}(x) + U(x) *
A_\mu * U^{-1}(x) \label{gt}
\end{equation}

Given a fermion field $\psi$, one can easily see that the
combination
\begin{equation}
\gamma^\mu D_\mu[A] \psi = \gamma^\mu \partial_\mu \psi - i e
\gamma^\mu A_\mu * \psi \label{p}
\end{equation}
transforms covariantly under gauge transformations (\ref{gt}),
\begin{equation}
\gamma^\mu D_\mu[A^U] \psi^U = U *\gamma^\mu D_\mu[A] \psi
\label{cov}
\end{equation}
with
\begin{equation}
\psi^U = U(x) * \psi \label{pp}
\end{equation}
and
\begin{equation}
\label{id} U(x) * U^{-1}(x) = U^{-1} * U(x) = 1
\end{equation}
A gauge invariant Dirac action can be defined in the form
\begin{equation}
S_f = \int d^dx\; \bar  \psi(x) * i\gamma^\mu D_\mu[A]\psi(x)
\label{Da}
\end{equation}

\subsection*{The Anomaly}
Chiral transformations will be written as
\begin{equation}
\psi'(x) = U_5(x)* \psi \label{chi}
\end{equation}
with
\begin{equation}
U_5(x) = \exp_*( \gamma_5 \alpha(x)) = 1 + \gamma_5 \alpha +
\frac{1}{2} \alpha(x)*\alpha(x) + \ldots
\end{equation}

The chiral anomaly ${\cal A}_d$ in $d$-dimensional space can be
calculated from the formula
\begin{equation}
\log J_d[\alpha]= -2  {\bf A}_d\; ,
\end{equation}
\begin{equation}
{\bf A}_d = \left.{\rm Tr} \; \gamma_5 \delta \alpha(x)
\right\vert_{reg} \label{reg}
\end{equation}
here $J_d[\alpha]$ is the Fujikawa Jacobian associated with an
infinitesimal chiral transformation $U = 1 + \gamma_5 \delta
\alpha$ and Tr includes a matrix and functional space trace.

Let us specialize to the two dimensional case. We shall use the
heat-kernel regularization so that (\ref{reg}) will be understood
as
\begin{equation}
{\bf A}_2=\int d^2x\; {\cal A}_2(x) * \delta\alpha(x)\;,
\label{reg1}
\end{equation}
\begin{equation}
{\cal A}_2(x) = \lim_{M \to \infty}{\rm Tr}\;  \gamma_5
\exp_*\left(\frac{\not \!\!D * \not \!\!D}{M^2}\right)\; .
\label{reg2}
\end{equation}
After some standard manipulations, (\ref{reg2}) takes the form
\begin{equation}
{\cal A}_2(x)  = \frac{1}{4\pi} {\rm tr}\; \gamma_5 {\not \!\! D} *
{\not \!\! D} = \frac{1}{4\pi} {\rm tr }\left(  \gamma_5 \gamma^\mu
\gamma^\nu \right)D_\mu* D_\nu \; .\label{ano}
\end{equation}
Here tr is  just the matrix trace. Using ${\rm tr}( \gamma_5
\gamma^\mu \gamma^\nu) = 2i\; \varepsilon^{\mu \nu}$,
 eq.(\ref{ano}) can be written as
\begin{equation}
{\cal A}_2(x)  =  \frac{e}{2\pi}
\varepsilon^{\mu\nu}(\partial_\mu A_\nu -ie  A_\mu* A_\nu) =
\frac{e}{4\pi} \varepsilon^{\mu\nu} F_{\mu\nu}\; .
\end{equation}
This result coincides with that first obtained in \cite{as}.

\section{The two-dimensional fermion determinant}

Let us write the gauge field in the two-dimensional case in the
form
\be
\as = \frac{1}{e} \left(\id\exp_* \left(\gamma_5\phi + i \eta
\right) \right)
* \exp_* \left(-  \gamma_5\phi - i \eta)
\right) \label{eq} \ee
Note that in the $\theta_{\mu \nu} \to 0$ limit, eq.(\ref{eq})
reduces to the usual decomposition of a two-dimensional gauge
field in the form
\be
e A_\mu =  \varepsilon_{\mu \nu} \partial^\nu \phi +
\partial_\mu \eta
\ee
which allows to decouple fermions from the gauge-field and then
obtain the fermion determinant as the Jacobian associated to this
decoupling \cite{rs}. Now, the form (\ref{eq}) was precisely
proposed in \cite{gss} to achieve the decoupling in the case of
non-Abelian gauge field backgrounds,  this leading to the
calculation  of the $QCD_2$ fermion determinant in a closed form.
Afterwards \cite{rf}, it was shown that writing a two dimensional
gauge field as in eq.(\ref{eq}) (without the $*$-product but in
the $U(N)$ case) does correspond to the choice of a gauge
condition  . Eq.(\ref{eq}) is then the  extension of this approach
for a case in which non-commutativity arises from the use of the
$*$-product.

At the classical level, the change of fermionic variables
\ba \psi = \exp_*\left( \gamma_5\phi + i \eta \right) * \chi
\nonumber\\ \bar \psi = \bar \chi * \exp_*\left( \gamma_5\phi - i
\eta \right) \label{cha} \ea
completely decouples the gauge field, written as in (\ref{eq}),
leading to an action of free massless fermions,
\be
S_f = \int d^2x\; \bar \chi * \id \chi \label{fa} \ee
Of course, this is not the whole story: at the quantum level there
is a Fujikawa Jacobian $J$ \cite{f} associated to change
(\ref{cha}). In order to compute this Jacobian, we follow the
method introduced in \cite{rs}-\cite{gss}. Consider then the
change of variables
\ba
\psi = U_t * \chi_t \; ,\nonumber\\
\bar \psi = \bar \chi_t *
U_t^\dagger \label{chat}
\ea
where
\be
U_t = \exp_*\left(t\left( \gamma_5\phi + i \eta \right)\right)\;,
\label{chatt} \ee
and $t$ is a real parameter, $0 \leq t \leq 1$. Given the fermion
determinant defined as
\be
\det ( \not \! \partial  -i e\as ) = \int \D\bar\psi \D\psi \exp\left( -S_f[\bar
\psi,\psi] \right) \label{Det} \ee
we proceed to the change of variables (\ref{chat}) which leads to
\ba
\det ( \not \! \partial -i e\as ) &=& J[\phi,\eta;t] \int \D\bar\chi_t
\D\chi_t \exp\left( -S_f[\bar \chi_t,\chi_t] \right) \nonumber\\
&= &J[\phi,\eta;t] \det D_t \label{DD}
\ea
where $J[\phi,\eta;t]$ stands for the Jacobian
\be
\D\bar\psi \D\psi= J[\phi,\eta;t] \D\bar\chi_t \D\chi_t
\label{jaco} \ee
and we have defined
\be
D_t = U_t^\dagger * (\not \! \partial -i e \as*)\;  U_t \label{dt} \ee
Now, since the l.h.s. in (\ref{DD}) does not depend on $t$ we get,
after differentiation,
\be
\frac{d}{d t}\log \det D_t = - \frac{d}{dt} \log J[\phi,\eta;t] \ee
or, after integrating on $t$ and using that $D_0 = \not \! \partial
 -i e\as$ and
$D_1 = \not \! \partial$
\be
\det (\not \! \partial -i e\as) = \det \not \! \partial\; \exp\left(-2\int_0^1 dt {\bf
A}_2(t)\right) \label{casi} \ee
where we have used
\be
{\bf A}_2(t) = \frac{d}{dt} \log J[\phi,\eta;t] \ee
Now, it is trivial to identify ${\bf A}_2(t)$ with the
two-dimensional chiral anomaly as defined in eq.(\ref{reg}), just
by writing $\delta \alpha = \phi dt$,
\be
{\bf A}_2 (t)=
\left. {\rm Tr}\left( \gamma_5*\phi \right) \right\vert_{reg} \ee
In order to have a gauge-invariant regularization ensuring that
the $\eta$ part of the transformation does not generate a
Jacobian, we adopt, in agreement with (\ref{reg1}) and
(\ref{reg2}),
\begin{equation}
{\bf A}_2(t) = \lim_{M \to \infty}{\rm Tr} \left( \gamma_5
\exp\left(\frac{\not \!\!D_t * \not \!\!D_t}{M^2}\right) *\phi
\right) \label{reg3}
\end{equation}
so that finally one has
\be
{\bf A}_2(t) = \frac{ e}{2\pi} \int d^2x \varepsilon^{\mu\nu}
\left(
\partial_\mu A_\nu^t -i e A_\mu^t * A_\nu^t\right) * \phi =
  \frac{ e}{4\pi}
\int d^2x\varepsilon^{\mu\nu} F_{\mu\nu}^t * \phi \ee
where we have introduced
\be
\gamma_\mu A_\mu^t= -\frac{1}{e} \left ( \id U_t \right)*U_t^{-1}
\ee
and analogously for $F_{\mu\nu}^t$. In summary, we can write for
the $U(1)$ fermion determinant
\be
\det (\not \! \partial -i e\as) =  \exp\left(-\frac{e}{2\pi}
\int d^2x\int_0^1 dt
\;\varepsilon^{\mu\nu} F_{\mu\nu}^t * \phi  \right)  \det \not \! \partial
\label{yasi} \ee

It will be convenient to use the relation
\be
\gamma^\mu \gamma_5 = -i \varepsilon^{\mu \nu} \gamma_\nu
\ee
to rewrite (\ref{yasi}) in the form
\be
\det (\not \! \partial  -i e\as) =  \exp\left(\frac{i e}{2\pi} {\rm tr}
\int
d^2x\int_0^1 dt \gamma_5 \phi * \left( \not \! \partial  \as^t
-i e \as^t * \as^t
\right) \right)  \det \not \! \partial
\label{yasis} \ee
Then, one can exploit the identity
\ba {\rm tr}\int d^2x
\frac{1}{2} \frac{d}{dt}  \as^t * \as^t &=& \frac{1}{e}
{\rm tr} \int d^2x
\gamma_5 \id \phi * \as^t + 2 {\rm tr} \int d^2x \gamma_5 \as^t *
\phi * \as^t \nonumber\\ &+& \frac{1}{e}{\rm tr}
\int d^2x (\not \!\! D\eta) * \as
\ea
and find for  (\ref{yasis})
\ba \log \det (\not \! \partial -i e\as) &=&  -\frac{e^2}{4\pi} {\rm tr}\int
d^2x\; \as *\as  +\frac{e^2}{2\pi}{\rm tr}
\int dt \int d^2x\; \gamma_5 \phi*
\as * \as  \nonumber\\ &+& \frac{e}{2\pi} \int dt
\int d^2x\;  (\not \!\! D\eta) * \as + \log \det \not \! \partial
\label{yasiss}
\ea

This is the final form for the fermion determinant in a $U(1)$
gauge theory. In order to write it in a more suggestive way
connecting it with the Wess-Zumino-Witten term, let us consider
the light-cone gauge $A_+ = 0$, Then, one can   see after some
algebra that \cite{r}
\ba
\log && \!\!\!\!\!\!\!\!
\left( \frac{\det (\not \! \partial -i e\as)}{\det \not\! \partial}
\right) =
-\frac{1}{8\pi} \int d^2x
\left(\partial_\mu g(x)^{-1} \right)   * \left(\partial_\mu g(x)\right)
\nonumber\\
&& \hspace{-1.3 cm}+ \frac{i}{12\pi} \epsilon_{ijk}\int_B d^3y g(x,t)^{-1} *
\left(\partial_i g(x,t)\right)
* g(x,t)^{-1} * \left(\partial_j g(x,t)\right)g^{-1} *
\left(\partial_k g(x,t)\right)
\nonumber\\
\label{WZ}
\ea
here we have written  $A_- = (i/e) g(x)*\partial_-g^{-1}(x)$ with
$g(x) = \exp_*(2\phi(x))$, $g(x,t) = \exp_*(2\phi(x) t)$ and $d^3y
= d^2x dt$ so that the integral in the second line of
eq.(\ref{WZ}) runs over the three dimensional manifold $B$, which
in compactified Euclidean space can be identified with a ball with
boundary $S^2$. Index $i$ runs from $1$ to $3$. As in the ordinary
commutative case, because the determinant was computed in
Euclidean space, elements $g$ should be considered as belonging to
$U(1)_C$ (the complexified $U(1)$) \cite{r}-\cite{ns}.

So, we have found  for the two-dimensional non-commutative fermion
determinant that, even for a $U(1)$ gauge field background, a
Wess-Zumino-Witten term arises due to non-commutativity of the
$*$-product. Of course, in the $\theta^{\mu\nu} \to 0 $ limit in
which the $*$-product becomes the ordinary one, the $U(1)$ fermion
determinant contribution to the gauge field effective action
reduces to $(-1/2\pi)\int d^2x \phi \partial^\mu\partial_\mu
\phi$ which is nothing but the Schwinger determinant result
expressed in a gauge-invariant way.

The method we have employed has the advantage that it can be
trivially generalized to the case of a $U(N)$ gauge group. One has
just to take into account that in (\ref{eq}) one has
\be
\phi = \phi^a t^a \, , \;\;\;\; \eta = \eta^a t^a \ee
with $t^a$ the $U(N)$ generators. Then, as originally shown in
\cite{gss} for the commutative case, the fermion determinant can
be seen to be given by
\be
\det (\not \! \partial -i e\as) =  \exp\left(-\frac{e}{4\pi}{\rm tr}^c\int
d^2x\int_0^1 dt \varepsilon^{\mu\nu} F_{\mu\nu}^t * \phi  \right)
\det \not \! \partial  \label{yasisss}
\ee
where $\rm{tr}^c$ is a trace over the $U(N)$ algebra. Then,
following the same steps leading to (\ref{WZ}), one gets, in the
$U(N)$ case
\ba \log \left( \frac{\det (\not \! \partial  -i e\as)}{\det \not \! \partial}
\right) &=&
-\frac{1}{8\pi}{\rm tr}^c  \int d^2x \left(\partial_\mu g^{-1}\right)
* \left(\partial_\mu g\right)
\nonumber\\
&&\hspace{-2cm} + \frac{i}{12\pi} \epsilon_{ijk} {\rm tr}^c \int_B
d^3y g^{-1} * \left(\partial_i g\right)
* g^{-1} * \left(\partial_jg\right)g^{-1} * \left(\partial_k g\right)
\nonumber\\
\label{WZc}
\ea
where again, in the light-cone gauge we have written
\be
A_- = -\frac{i}{e} g * \partial_- g^{-1} \, , \;\;\; A_+ = 0 \ee
\be
g = \exp_*(2 \phi^a t^a) \ee
Eq. (\ref{WZc}) is the generalization of the expression given in
\cite{PW}  for the
two-dimensional non-Abelian fermion determinant to the case of
non commuta\-tive space-time.

%
%
\section{Conclusion}

We studied in this article the effective action of the gauge
degrees of freedom in a two dimensional non-commutative Field
Theory of fermions coupled to a gauge field. Using Fujikawa's
approach, we computed the chiral anomaly and, from it, the
fermionic determinant of the non-commutative Dirac operator.

As it was to be expected, the result for the fermion determinant
corresponds to the $*$-deformation  of the standard result. Now,
the fact that a Moyal bracket enters in the field strength
curvature even in the Abelian case, has important consequences,
some of which have already been signaled in \cite{as}-\cite{bst}
where chiral and gauge anomalies in non-commutative spaces have
been analyzed.

In our framework, where the anomaly was integrated in order to
obtain the fermion determinant, this reflects in the fact that a
Wess-Zumino-Witten like term arises \underline{both} in the
Abelian and in the non-Abelian cases (eqs.(\ref{WZ}) and
(\ref{WZc}) respectively). This should have, necessarily,
implications in relevant aspects of two-dimensional theories
since, as it is well-known, bosonization is closely related to the
form of the fermion determinant \cite{dVR}. Indeed, the
bosonization rules for fermion currents as well as the resulting
current algebra can be easily derived by differentiation of the
Dirac operator determinant $\det(\!\!\not\!\! d -i\!\!\!\not\!\!
s)$ with respect to the source $s_\mu$ (see \cite{gmns} for a
review). Now, as one learns from ordinary non-Abelian
bosonization, where the Polyakov-Wiegmann identity plays a central
r\^ole in the bosonization recipe, here one should have an
analogous identity which will lead to non-trivial changes  at the
level of currents and, a fortiori, for the current algebra. In
view of the relevance of these objects in connection with
two-dimensional bosonic and fermionic models,
 it will be worthwhile to
pursue the investigation initiated here in this direction.

~
\subsection*{Acknowledgements}
F.A.S. would like to thank Claude Viallet for discussions on
anomalies in non-commutative spaces and LPTHE, Paris U. VI-VII for
hospitality. This work is  supported in part by grants from CICBA,
CONICET (PIP 4330/96), and ANPCYT (PICT 97/2285) Argentina.


\end{document}